\documentclass[aps,prb,twocolumn,superscriptaaddress]{revtex4}
\usepackage{subfigure}
\usepackage{amsmath}
\usepackage{graphicx}
\usepackage{rotating}
\usepackage{amsfonts}
\usepackage{amssymb}
\begin{document}
\title{Tight-binding Hamiltonian for LaOFeAs}
\author{D.A. Papaconstantopoulos}
\affiliation{Department of Computational and Data Sciences, George
  Mason University, Fairfax VA 22030}
\author{M.J. Mehl}
\affiliation{Center for Computational Materials Science, Naval Research Laboratory, Washington DC 20375}
\author{M.D. Johannes}
\affiliation{Center for Computational Materials Science, Naval Research Laboratory, Washington, DC 20375}
\date{Printed on \today}

\begin{abstract} 
First-principles electronic structure calculations have been very
useful in understanding some of the properties of the new iron-based
superconductors. Further explorations of the role of the individual
atomic orbitals in explaining various aspects of research in these
materials, including experimental work, would benefit from the
availability of a tight-binding(TB) Hamiltonian that accurately
reproduces the first-principles band structure results. In this work
we have used the NRL-TB method to construct a TB Hamiltonian from
Linearized Augmented Plane Wave(LAPW) results. Our TB model includes
the Fe d-orbitals, and the p-orbitals from both As and O for the
prototype material LaOFeAs. The resulting TB band structure agrees
well with that of the LAPW calculations from 2.7 eV below to 0.8 eV
above the Fermi level, $\varepsilon_F$, and the Fermi surface
matches perfectly to that of the LAPW.  The TB densities of
states(DOS) are also in very good agreement with those from the LAPW
in the above energy range, including the per orbital
decomposition. We use our results to provide insights on the
existence of a pseudogap in the DOS just above the Fermi level.  We
have also performed a separate TB fit to a database of LAPW results
as a function of volume and with variations of the As positions.
This fit although less accurate regarding the band structure near
$\varepsilon_F$, reproduces the LAPW total energies very well and
has transferability to non-fitted energies.

\end{abstract}

\maketitle

\section{Introduction}

The discovery of superconductivity with critical temperature Tc of
about 55K in the iron compounds named iron-Oxypnictides
\cite{kamihara,ren} has brought to the field excitement comparable
to that created by the high Tc cuprates. The prototype formula for
these materials is LaOFeAs with two distinct layers of LaO and
FeAs. The room-temperature crystal structure is tetragonal and
undergoes a structural distortion to orthorhombic at low
temperatures \cite{cruz}. The transition temperature is modulated by
electron doping (F substituting for O)\cite{kamihara} or hole doping
(Sr substituting for La)\cite{wen}. Other substitutions may occur,
{\em e.g.}, replacing Fe by Ni \cite{cao} or As by
P\cite{kamihara_1}.

Many experimental and theoretical investigations have been performed
without reaching a consensus as to whether these materials are
similar to the cuprates, with their superconducting mechanism not
yet understood, or if the essential physics is different, indicating
that this is yet another class of superconductors.  Questions to be
answered include the nature of the normal state, the symmetry of the
superconducting state and, of course, the origin of the pairing
interaction.

Density functional theory (DFT)
calculations\cite{hohenberg64:dft,kohn65:inhom_elec} have been at
the center of the theoretical investigations. For LaOFeAs, Singh and
Du \cite{singh} found a high density of states at the Fermi level,
N($\varepsilon_F$), and a low carrier concentration. We note that
this is different from the cuprates which have a low
N($\varepsilon_F$) and while displaying low carrier concentrations
they are characterized by a half-filled band near $\varepsilon_F$.
On the other hand, Singh and Du found a high N($\varepsilon_F$) with
antiferromagnetic fluctuations, which has a definite similarity with
the cuprates. On the same theme, competing antiferromagnetism and
superconductivity in the doped system is suggested by Yildirim
\cite{yildirim} as breaking the tetragonal symmetry causing a
structural distortion.

Since standard DFT is very expensive computationally there is a need
for developing tight-binding (TB) models that can be the starting
point for carrying out further investigations using many-body
techniques, such as multiband Hubbard models.  Having examined the
details of the DFT calculations for LaOFeAs, we identified the
following features of the energy bands emerging from a wave function
analysis. Starting from the lower bands we find O p-states that, for
higher energies, hybridize with As p-states. Hybridization with Fe
d-states occurs at the top of the As p-states and then hybridization
with La states appears high above the Fermi level. Therefore, it
becomes clear that TB models that ignore the other elements and use
only the Fe d-orbitals are not reproducing the band structure of
LaOFeAs accurately enough.

Several TB approaches have appeared in the literature, such as the
study of Kuroki {\em et al.} \cite{kuroki}, which is based on an
Fe-only d-band Hamiltonian. These authors applied the random phase
approximation to obtain spin and charge susceptibilities, concluding
that an unconventional s-wave pairing is in play.  Furthermore, a
recent paper by Manousakis {\em et al.}  \cite{manousakis} builds a
TB Hamiltonian fitted to DFT results using in addition to the Fe
d-orbitals p orbitals of As. Based on their TB Hamiltonian these
authors report that the effective Hamiltonian, in the strong on-site
Coulomb-repulsion limit, operates on three distinct subspaces
coupled through Hund's rule. They also argue that the observed
spin-density-wave order minimizes the ground state energy of the
Hamiltonian. These conclusions could be a correct speculation of the
physics in this material. However, their calculations are based on a
TB Hamiltonian that is not accurately derived from the
first-principles data. Although they performed their TB fit only
near the Fermi level, their results do not reproduce the energy
bands well enough, as can be seen in their Fig~4. Therefore, the
conclusions of this paper, based on a poorly constructed TB
Hamiltonian, can only be considered as a speculation. More recently
a paper by Eschrig and Koepernik \cite{eschrig} presents a minimal
basis TB Hamiltonian for LaOFeAs, as well as for other structure
types of the Fe superconductors, that is based on an elegant TB
theory. However, this Hamiltonian has quantitative agreement to
first-principles results only in a smaller window around E$_F$ than
the fit we present here.

There are two other 5-orbital tight-binding studies with which we
compare in our results section.  These are the work of
Calder\'{o}n {\em et al.}\cite{calderon09:tbfepnictides} and Graser
{\em et al.}.\cite{graser09:pnictide}

In this work we have used the Naval Research Laboratory
Tight-Binding (NRL-TB) \cite{mehl,johannes} method to fit our
linearized augmented plane wave (LAPW) results
\cite{koelling77,wei85:lapw,singh86:_accel}to a TB basis with the
aim of reproducing the band structure very accurately. We have
included the d orbitals of Fe, the p orbitals of As, and the p
orbitals of O, leaving out the La orbitals since their effect is
only evident high above $\varepsilon_F$. In this study we examine
the effect of each of the above orbitals on how accurately the
first-principles band structure can be reproduced.

In this work we fit the NRL-TB method to an LAPW band structure
ranging from 2.7 eV below to 0.8 eV above $\varepsilon_F$.  The TB
band structure fits the LAPW results very well near $\varepsilon_F$
and perfectly reproduces the Fermi surface near the $\Gamma$ and $M$
symmetry points.

The TB densities of states are also in very good agreement with the
corresponding LAPW results, including a comparison by orbital
decomposition. In addition, we have studied the variation of the
total energy with respect to the position of the As atoms. We have
found that the TB total energies fit the LAPW values very well even
for energies that we did not include in our fit.

\section{Computational details}

The equilibrium structure of non-magnetic LaOFeAs is the AsCuSiZr
structure \cite{johnson}.  This structure is tetragonal, space group
P4/nmm, with eight atoms in the unit cell and lattice constants
$a = 7.626$~Bohr and $c = 16.518$~Bohr.  The Oxygen atoms occupy the
(2a) Wyckoff positions, and Iron the (2b) positions.  The La and As
atoms occupy (2c) sites, with $z = 0.14154$ for La and $0.6512$ for
As.  We used a regular, $\Gamma$-centered $8 \times 8 \times 4$
k-point mesh, which results in 225 points in the irreducible part of
the Brillouin zone.  The LAPW basis functions were cut off at
$RK_{max} = 8.5$, with approximately 1250 basis functions at each
k-point.  To help convergence we broadened the spectrum using a
Fermi distribution at a temperature of 5~mRy.

As expected our LAPW energy bands and densities of states are
basically identical to those published by Singh and Du \cite{singh}.
We also performed 21 additional LAPW calculations by varying the
above structural parameters $a$ and $z_{\mbox{As}}$ for the purpose
of creating a first-principles total energy database to use in our
TB calculations.

Our TB Hamiltonian was built following the NRL-TB method. We
summarize below the basic equations of this scheme which is based on
a Slater-Koster approach\cite{slater54:tb} with two-center
parameters.


Unlike the general NRL-TB method, where we include $s$, $p$, and $d$
orbitals for each atom, here our basis set includes only the $d$
orbitals for Fe and the $p$ orbitals for As and O.  All other
contributions to the band structure, including the effects of La,
are ignored as they have little weight in the region between the
bottom of the As-O $p$ bands and until well above the Fermi level.
Since there are two Fe, As, and O in each unit cell of the
structure, we end with a 22$\times$22 matrix to diagonalize at
each k-point.  In addition, we limit ourselves to an orthogonal
Hamiltonian, so we ignore the possible overlap hopping parameters.

{\em Onsite Parameters} In the NRL-TB, the onsite energies are
determined by the interaction of an individual atom with its
environment.  In our study of LaOFeAs, however, we are considering
at most small displacements of the atoms around the equilibrium
positions.  We therefore use a constant value for the onsite
parameters.  However, we recognize that the symmetry of the iron
states allows the orbitals to have different onsite parameters.
Accordingly, we chose four onsite parameters for Fe:  one for the
$xz$ orbital, which will be equal to the $yz$ parameter, and one each
for the $xy$, $x^2-y^2$ and $3z^2-r^2$ orbitals.  We also have one
onsite parameter for the As $p$ orbitals, and another for the O
$p$ orbitals, for a total of six independent onsite parameters,
listed in Table \ref{tab:onsite}.

{\em Hopping Parameters} The usual $spd$ Slater-Koster scheme has 10
two-center hopping parameters for like-atom interactions and 14
parameters for the interaction of two unlike atoms.  Since we limit
our system to Fe $d$ and As/O $p$ orbitals, we have only the
following interactions:

Fe-Fe:  $dd\sigma$, $dd\pi$, $dd\delta$

As-As (or O-O): $pp\sigma$, $pp\pi$

Fe-As (or Fe-O) (Fe atom first):  $dp\sigma$, $dp\pi$

As-O: $pp\sigma$, $pp\pi$

\noindent for a total of 13 Slater-Koster parameters for each
neighbor distance.  As with the standard NRL-TB, these parameters
depend only on the distance, $R$, between the two atoms, and are
parametrized according to the formula

\begin{equation}
H_{\ell \ell' \mu} (R) = (a_{\ell \ell' \mu} + b_{\ell \ell' \mu}
R + c_{\ell \ell' \mu} R^2) \exp (-d_{\ell \ell' \mu}^2 R) F(R) ~,
\label{equ:hopping}
\end{equation}
where $a, b, c$ and $d$ are fitting parameters, and the cutoff
function $F(R)$ has the form
\begin{eqnarray}
F(R) & = & 1/\{1 + \exp[(R - R_0)/\ell]\} , R < R_0 \nonumber \\
& = & 0, R > R_0 ~ .
\label{equ:cutoff}
\end{eqnarray}
In our fits we take $R = 14$~Bohr and $\ell = 0.5$.


The parameters in the above equations are determined by a
least-squares fit to the LAPW eigenvalues at 225 k-points in the
irreducible Brillouin zone. We have found that applying group theory
to block-diagonalize the Hamiltonian at high symmetry points is
essential for obtaining a good fit.

\section{Results}

The onsite parameters determined from our fitting procedure are
given in Table~\ref{tab:onsite}.  To make our results easier to
compare to first-principles calculations we have rotated the
Cartesian axis by 45$^\circ$ relative to the primitive vectors of
the tetragonal unit cell, so that the lobes of the $x^2-y^2$ Fe
orbitals point to the nearest-neighbor Fe atoms.

\begin{table} 
\caption{Onsite parameters for Fe-As-O, determined using the methods
  described in the text.  All energies are in Rydbergs.}
\begin{tabular}{ccr}
Atom & Orbital & Onsite \\ \hline
Fe & $yz$ ($zx$) & 0.51108 \\
& $xy$        & 0.54617 \\
& $x^2-y^2$   & 0.54548 \\
& $3z^2-r^2$  & 0.5513 \\ \hline
As & $p$         & 0.18566 \\ \hline
O  & $p$         & 0.39230
\end{tabular}
\label{tab:onsite}
\end{table}

In Table~\ref{tab:tbpar} we list the Slater-Koster hopping
parameters generated by Eq. \ref{equ:hopping}.

\begin{table}
\caption{Slater-Koster hopping parameters generated by
  Eq.~(\ref{equ:hopping}), as described in the text. Energy units
  are in Rydbergs, and distances are in Bohr.}
\begin{tabular}{cccccc}
Fe & Fe & R &  $dd\sigma$ & $dd\pi$  & $dd\delta$ \\  
&        & 5.329& -0.02771 & 0.01001 & 0.00031 \\
&        & 7.626 & 0.00546 & 0.00029 & 0.00750 \\
&        & 10.784 & 0.00364 & -0.00500 & 0.00008 \\ \hline
As & As & R & $pp\sigma$ & $pp\pi$ & \\
&    &  7.350 & 0.05880 & 0.08276 &  \\
&    &  7.626    & 0.06633 & 0.04262 &  \\ 
&    &  10.784  & 0.01041  & -0.05779 & \\ \hline
O  & O  & R & $pp\sigma$  & $pp\pi$ & \\ 
&    & 5.392  & 0.01885 & -0.00783 &  \\
&    & 7.626 & 0.00939 & -0.00534 & \\ 
&    & 10.784 & 0.00208 & 0.00085 & \\ \hline
Fe & As & R & $dp\sigma$ & $dp\pi$ & \\
&    & 4.558  & 0.17916 & 0.00931 &  \\
&    & 8.884 & -0.00751 & -0.02974 &  \\
&    & 11.708 & -0.00073 & -0.00090 & \\ \hline
Fe & O  & R & $dp\sigma$ &  $dp\pi$ & \\
&    & 8.259 & -0.00319 & -0.00338 &  \\
&    & 9.863 & 0.00021 & 0.00240 & \\ 
&    & 11.241 & -0.01449 & 0.00648 & \\ \hline
As & O  & R & $pp\sigma$ & $pp\pi$ & \\
&    & 6.909 &0.00513 & -0.02238 & \\
&    & 10.290 & 0.01562 & -0.00206 & \\
&    & 11.412 & 0.00591 & -0.00028 & \\ \hline
\end{tabular}
\label{tab:tbpar}
\end{table}

\begin{figure}
\label{fig:13bands}
\includegraphics[width = 0.95\linewidth]{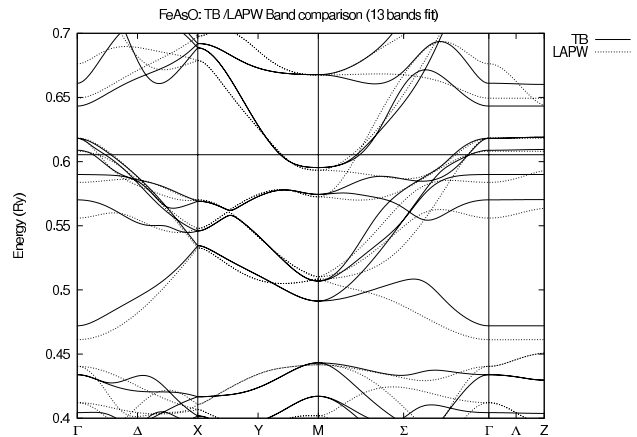}
\caption{A comparison of the LAPW (solid line) and tight-binding
  (dashed line) band structure for the bands near the Fermi level.
  The horizontal line at $y = 0.605$~Ry is the LAPW Fermi level.}
\label{bands}
\end{figure}

These parameters are designed to fit the 9$^{th}$ to 21$^{st}$ bands
of our TB Hamiltonian to first principles results for three nearest
neighbors of Fe-Fe, As-As, O-O, Fe-As, Fe-O, and As-O interaction.In
Table~\ref{tab:bandrms} we show the RMS error per band for this fit,
indicating the high quality of the fit to the LAPW eigenvalues. It
should be noted that near the Fermi level which is in the vicinity
of the 17$^{th}$ and 18$^{th}$ bands the deviation between LAPW and
TB is on the average, including all the 225 k-points, 3mRy. At the
high symmetry points such as at $\Gamma$, $X$ and $M$ the deviation
is less than 1mRy, as can be seen in Fig.~\ref{bands} which shows a
comparison between LAPW and TB results.

The other TB papers
\cite{kuroki,manousakis,eschrig,calderon09:tbfepnictides,graser09:pnictide}
do not report the detailed quantitative information contained in our
Table~\ref{tab:bandrms}, which makes it difficult to have a detailed
comparison of TB models.

\begin{table}
\caption{RMS ERROR: per band for 225 k-points}
\begin{tabular}{rr}
Band & RMS Error (Ry) \\ \hline
9   &  0.008365 \\    
10   &  0.009243 \\    
11   &  0.011206 \\     
12   &  0.009935 \\     
13   &  0.011851 \\     
14   &  0.007552 \\     
15   &  0.005671 \\     
16   &  0.005491 \\     
17   &  0.002917 \\     
18   &  0.002912 \\     
19   &  0.007435 \\     
20   &  0.008658 \\    
21   &  0.012827 \\     
22   &  0.010253 \\     
\end{tabular}
\label{tab:bandrms}
\end{table}

The aforementioned 13 bands span an energy range from 0.4Ry to
0.75Ry as shown in Fig.~\ref{bands}. In this figure one can see that
from 0.45 Ry where a gap is present to 0.70 Ry the fit is good and
around the Fermi level(0.605Ry) the agreement between TB and LAPW is
excellent; the energy bands clearly show holes around the center of
the Brillouin zone and electron pockets around the high symmetry
point M. This suggests that the TB Fermi surface preserves all the
characteristic features found in the first-principles
calculations. 

Our model also reproduces the LAPW Dirac cone, slightly shifted to
the right and higher in energy by approximately 12.5~mRy
(170~meV). There is a secondary erroneous crossing in the vicinity
that is confusing to the eye.  The TB model of Calder\'{o}n {\em et
  al.}  \cite{calderon09:tbfepnictides} puts the Dirac cone at least
1~eV higher while Kuroki {\em et al.} \cite{kuroki} place it even
higher. The model of Maier {\em et al.} \cite{graser09:pnictide}
seems to capture the Dirac cone (but may miss a Fermi crossing at
$\Gamma$) as does that of Eschrig {\em et al.} \cite{eschrig},
though this is not a five-band model, but rather a downfolding
technique which is in a different spirit than our approach.  Of
course none of the five-band models (nor any effective Fe-Fe only
model) has the capability to turn individual (as compared to
effective) hopping parameters on and off, as we do in a later
section to understand the origin of the pseudogap.

\begin{figure}
\includegraphics[width=0.85\linewidth, angle=270]{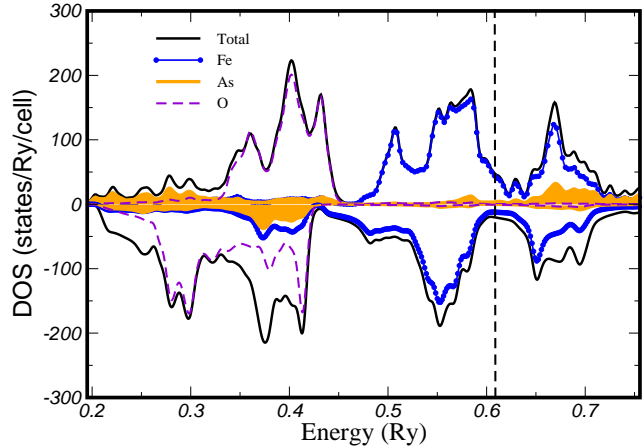}
\caption{(color online) Comparison of the LAPW (lower half) and TB
  (upper half) DOS.  Site decomposition for Fe, As, and O is also
  shown.}
\label{site}
\end{figure}

In Fig.~\ref{site}, we show a comparison between TB
and LAPW densities of states.  In Fig.~\ref{site} the total DOS are
compared where one can see that TB matches well the two LAPW peaks
below $\varepsilon_F$ and the one peak above $\varepsilon_F$. The
site decomposed DOS is also shown in Fig.~\ref{site} and shows very
good agreement for the Fe d-states, and reasonable to good agreement
for the As and O $p$-states. One can conclude that the TB produces
reliable results not only for the eigenvalue spectrum but also for
the eigenvectors.

The orbitally decomposed DOS are also in at least semi-quantitative
agreement between LAPW and TB results. Note that since the LAPW
angular-momentum components of the DOS are projections inside the
muffin-tin spheres, an exact comparison cannot be made.  Still, the
TB is a powerful tool in explaining features of the DOS that cannot
be addressed by the first-principles calculations. For example, we
can trace down the origin of the the so-called pseudogap above
$\varepsilon_F$ shown in Fig.~\ref{site}.  Since the local
environment of the Fe atom is tetragonal (distorted to some degree
in most 1111 type compounds), and given that the strongest hopping
parameter is the Fe-As $dp\sigma$, as seen in Table~\ref{tab:tbpar},
one might expect to see a gap or pseudogap between a lower $e_g$
doublet and an upper $t_{2g}$ triplet, commensurate with the local
symmetry.  However, in reality the gap occurs between a lower peak
in the DOS containing three states per Fe (six total) and and upper
peak containing two states per Fe (four total), $i.e.$ the reverse
of the naive ligand field (or crystal field) expectation (this can
be seen in Fig.~\ref{site} in the Fe part of the spectrum).  Here we
use our TB model to eliminate hoppings one by one and trace down the
origin of the pseudogap.

\begin{figure}
\includegraphics[width = 0.85\linewidth, angle=270]{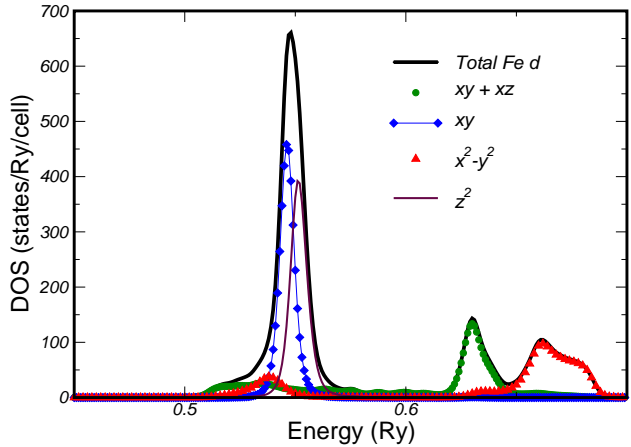}

\caption{(color online) The DOS with only nearest neighbor Fe-As
  hoppings included.  The expected ligand field splitting
  corresponding to the tetragonal environment of the Fe atom is
  clearly visible.}

\label{ligandfield}
\end{figure}

First we turn off all but the Fe-As nearest neighbor hoppings and
indeed find a lower doublet and upper triplet, as must be the case
(see Fig.~\ref{ligandfield}).  The splitting between triplet
states, and to a lesser degree the doublet states, is due to the
imperfect tetrahedron.

\begin{figure}
\includegraphics[width = 0.80\linewidth, angle = 270]{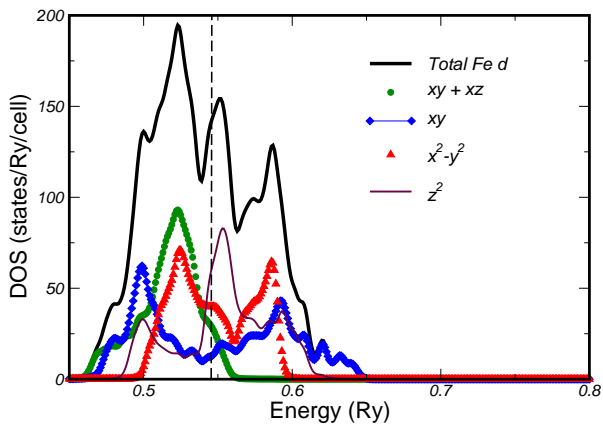}
\caption{(color online) The DOS with all hoppings eliminated except
  direct hopping between Fe atoms.  Most of the orbitals split into
  bonding/anti-bonding combinations with only the $xz+yz$
  combination remaining as a single peak.}
\label{FeFe}
\end{figure}

Next, we eliminate all Fe-As hopping, but allow hopping between the
two distinct Fe (these are nearest Fe neighbors) in the unit cell.
As can be seen in Fig.~\ref{FeFe}, this has the effect of creating
bonding/anti-bonding complexes within each orbital designation.
Interestingly, the $xz+yz$ orbitals do not split at all, remaining
as a single peak, while the most strongly split orbital is $xy$.
Already, something like a pseudogap can be seen to be forming around
0.57 Ry, with a spectral distribution quite different from the
ligand field gap distribution in Fig.~\ref{ligandfield}.

\begin{figure} 
\includegraphics[width = 0.80\linewidth, angle=270]{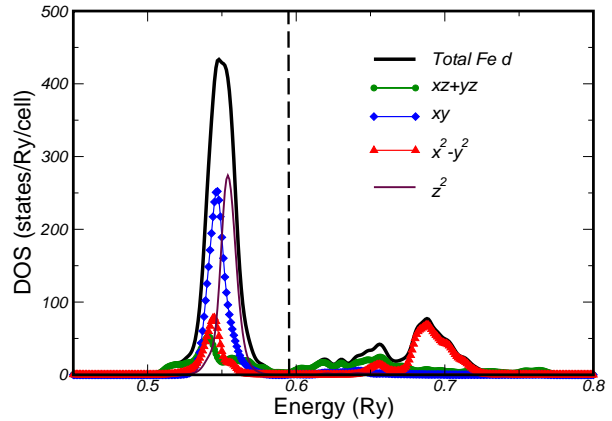} 
\caption{(color online) The DOS including all Fe-As hoppings, but
  without any direct Fe-Fe hopping.  The $z^2$ and $xy$ orbitals
  remain unsplit, while all other orbitals show rather strong
  bonding/anti-bonding peak separations.}
\label{FeAsall} 
\end{figure}

Finally, we allow {\em all} Fe-As hoppings, but remove direct Fe-Fe
hopping.  This scenario accounts for interaction between the two Fe
sites in the unit cell via As.  In Fig.~\ref{FeAsall}, a
bonding/anti-bonding splitting again occurs in most of the orbitals
with the strength of the splitting considerably stronger than that
initiated by Fe-Fe direct hopping.  In this case the $xy+xz$ orbital
does undergo splitting, but the $z^2$ and $xy$ orbitals do not.  A
fairly strong pseudogap emerges with the majority of the weight in
the lower complex.  Note that there are still significant
differences between the DOS in Fig. \ref{FeAsall} and the full TB
model DOS in Fig. \ref{site}.  This underlines the importance of
direct Fe-Fe hopping.

\begin{figure} 
\includegraphics[width = 0.85\linewidth]{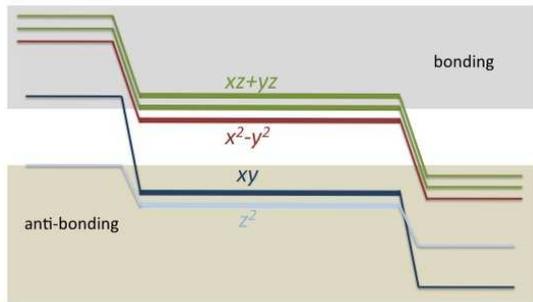} 

\caption{(color online) A schematic illustrating the origin of the
  pseudogap.  Each of the original five orbitals is doubly
  degenerate (one set coming from each Fe in the unit cell). The
  $xz+yz$, $x^2-y^2$, and $xy$ orbitals bonding and anti-bond with
  sufficient strength to form upper and lower complexes, designated
  by the two shaded regions.  The $z^2$ orbital has a small
  splitting, but both peaks remain in the lower complex.  The
  pseudogap and Fermi energy lie between the upper (anti-bonding
  only) and lower (mostly bonding) complexes.}

\label{cartoon} 
\end{figure}

Combining the information from all three reduced hopping diagrams,
we can understand how the pseudogap forms. It is the result of a
combination of strong and weak bonding/anti-bonding splitting of the
orbitals, due to the two distinct Fe atoms in the unit cell. Hopping
occurs both via As and directly between Fe atoms (further hoppings
also surely contribute somewhat).  In Fig.~\ref{cartoon}, we show
how very strong splitting of the $t_{2g}$-triplet derived states
coupled with weaker splitting of the $e_g$ doublet-derived states
results in the calculated pseudogap.  The $xz+yz$, $x^2-y^2$
orbitals split strongly into two peaks per orbital, forming an upper
complex of three states and a lower one also of three states.  The
$xy$ orbital has a weaker splitting, but strong enough still to
place one state in the upper complex and one in the lower.  Not
surprisingly, the $z^2$ orbital, which is pointed mainly
out-of-plane, has the weakest splitting such that both the bonding
and anti-bonding peaks remain in the lower complex.  Thus, there are
six states in the lower complex and four in the upper.  Note that
the actual splitting of the $xy$ state is somewhat more complicated
than the simple bonding/anti-bonding schematic suggests.  This
simply mirrors the fact that the DOS itself is not composed of two
simple peaks of precisely six and four states each, but rather has
some secondary peak structures.  In addition, all of the states are
broadened into bands which, along with strict symmetry requirements
\cite{eschrig} cause finite overlap between the upper and lower
complexes and give rise to a pseudogap rather than a full gap. By
turning off specific hoppings, we have elucidated from an atomic
orbital point of view, the mechanism that gives rise to the
pseudogap.  It requires both direct Fe-Fe interaction {\em and}
interaction via the intermediate As.

\begin{figure}
\includegraphics[width = \linewidth]{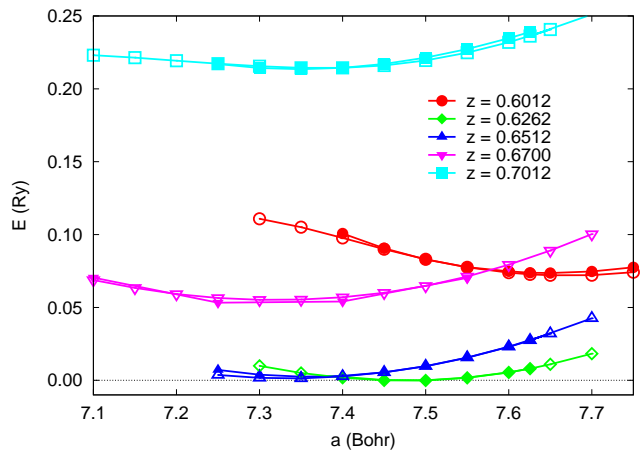}
\caption{(color online) The total energy curves generated by TB fits
  (open symbols) to LAPW data (solid symbols) for various values of
  the As height above the Fe plane and planar lattice constant, for
  fixed lattice constant c = 16.4~$a.u.$}
\label{alleng}
\end{figure}

The second fit that we performed has the objective to fit the volume
and the As position variations of the total energy. The height of
the As ion has strong effects on the electronic structure and
magnetism and may even be able to switch the pairing symmetry
\cite{pickett,mazin,kuroki}. For this purpose we run 21 separate
LAPW calculations to fit in our TB scheme the total energy. In this
calculation we fitted all the 22 energy bands that correspond to our
Fe(d)-As(p)-O(p) TB Hamiltonian. This fit does not give the excellent
fit to the LAPW Fermi surface that we found in the first
fit. However, it fits the 21 LAPW total energies perfectly with an
RMS error of 0.0004 Ry. In Fig.~\ref{alleng} we show the total
energy results of this fit. We used these TB parameters to calculate
total energies outside our database and compared with independent
LAPW results not included in the fit and found very good
agreement. Using these results we found the lattice equilibrium
parameters to be a=7.4 $a.u.$, c=16.4 $a.u$ and As position z=0.64 (giving 113.5$^\circ$ as the As-Fe-As angle),
in reasonable agreement with the experimental values.

\section{Conclusions}

We report TB results on LaOFeAs obtained by the NRL-TB method via a
fit to LAPW eigenvalues and total energies. Two TB parametrizations
were performed: the first aims at reproducing the energy bands in an
energy range from 2.7~eV below to 0.8~eV above $\varepsilon_F$ with
superior accuracy around $\varepsilon_F$. From this parametrization
an analysis of the orbital-decomposed DOS shows that the mechanism
which creates the pseudogap above $\varepsilon_F$ comes from a
direct Fe-Fe interaction and from an Fe-Fe interaction through the
intermediate As atom. In our second TB parametrization we focus on
the energetics of the LaOFeAs system finding a TB Hamiltonian that
fits the LAPW total energies as a function of volume and As position
very well. We propose that this TB Hamiltonian will be very useful
in carrying out many-body theory with a more realistic Hamiltonian
than those employed previously containing just the d-iron orbitals.



\section*{Acknowledgments}

We thank Igor Mazin for useful discussion. The work at GMU is
supported by ONR grant No: N000140911025, and the work at NRL is
supported by the Office of Naval Research.
 

\end{document}